\documentclass[12pt]{article}
\pdfoutput=1  %needed for JHEP at least?
\usepackage{hyperref}
\usepackage[pdftex]{graphicx}
\usepackage{amsmath,amssymb}
\usepackage{cancel}
\usepackage{appendix}
\newcommand{\gs}{\ensuremath{g_s}} % String coupling constant
\newcommand{\ls}{\ensuremath{l_s}} % String length
\def\p{\partial}

\newcommand{\cL}{\mathcal{L}}

\newcommand{\cN}{{\mathcal{N}}}

\newcommand{\bS}{{\mathbf{S}}}

\newcommand{\vev}[1]{{\left< {#1} \right>}}

\newcommand{\be}{\begin{equation}}
\newcommand{\ee}{\end{equation}}
\newcommand{\bea}{\begin{eqnarray}}
\newcommand{\eea}{\end{eqnarray}}

\begin{document}

\begin{titlepage}

\begin{flushright}
PUPT-2471\\
TCC-020-14\\
UTTG-18-14
\end{flushright}

\begin{center}
\Large \bf \textbf{Branes from Light: Embeddings and Energetics for Symmetric $k$-Quarks in $\cN=4$ SYM}
\end{center}

\begin{center}
Bartomeu Fiol$^{\ast}$,
Alberto G\"uijosa$^{\diamond}$\footnote{Work done during a sabbatical year at the
Department of Physics at Princeton University.}
and Juan F.~Pedraza$^{\dagger,\ddagger}$

\vspace{0.2cm}
$^{\ast}$
Departament de F\'isica Fonamental i Institut de Ci\`encies del Cosmos,\\
Universitat de Barcelona,
Marti i Franqu\`es 1, 08028 Barcelona, Catalonia, Spain \\
\vspace{0.2cm}
$^{\diamond}\,$Departamento de F\'{\i}sica de Altas Energ\'{\i}as, Instituto de Ciencias Nucleares, \\
Universidad Nacional Aut\'onoma de M\'exico,
\\ Apartado Postal 70-543, M\'exico D.F. 04510, M\'exico\\
 \vspace{0.2cm}
$^{\dagger}\,$ Theory Group, Department of Physics and Texas Cosmology Center, \\
University of Texas, 1 University Station C1608, Austin, TX 78712, USA\\
\vspace{0.2cm}
$^{\ddagger}\,$
Perimeter Institute for Theoretical Physics, \\
Waterloo, Ontario N2L 2Y5, Canada \\
\vspace{0.2cm}
{\tt bfiol@ub.edu, alberto@nucleares.unam.mx, jpedraza@physics.utexas.edu}
\vspace{0.2cm}
\end{center}

\begin{center}
{\bf Abstract}
\end{center}
\noindent
We construct the D3-brane dual to a $k$-quark of $\cN=4$ super-Yang-Mills theory in the totally symmetric representation of $SU(N)$, undergoing \emph{arbitrary} motion. Our method of construction generalizes previous work by Mikhailov, and proceeds by shooting light rays inward from the anti-de Sitter boundary, to trace out the brane embedding. We expect this method to have wider relevance, and provide evidence for this by showing that it correctly reproduces the known D5-brane embeddings dual to totally antisymmetric $k$-quarks. As an application of our solutions, we compute the energy of the D3-brane and extract from it the $k$-quark's intrinsic energy and rate of radiation. The result matches expectations based on previous calculations, and makes contact with the exact Bremsstrahlung function for the fundamental representation.
 \vspace{0.4cm}
\smallskip
\end{titlepage}

\tableofcontents

\section{Introduction and Summary}\label{introsec}

\subsection{Motivation and antecedents}\label{motivationsubsec}

One of the useful ways to inquire into the behavior of a field theory is to couple an external probe to it and deduce the response of the fields. The vevs of line operators associated with such probes serve as diagnostics of the phase of the field theory. Furthermore, even for a simple theory such as Maxwell's,
%or Maxwell (analogous to how we use "Yang-Mills"), or Maxwell electrodynamics
once one considers probes with generic motion, one encounters the fascinating phenomenon of radiation, with far-reaching implications both on a theoretical and a practical level. For generic interacting quantum field theories, the analytic treatment of detailed properties of radiation, like its frequency and angular distribution, its broadening, or the damping it induces on the probe, is mostly based on perturbation theory and therefore limited to the realm of weak coupling. In recent years, it has been appreciated that for quantum field theories with additional symmetries ({\it e.g.}, conformal invariance and/or supersymmetry) and for suitably chosen probes, there are a number of techniques that allow us to address these questions for strongly-coupled field theories. In very fine-tuned examples, and for very specific questions, we can even obtain exact answers, something quite unusual for a field theory in more than two dimensions.

Depending on the questions we want to address, we sometimes need to be able to handle probes with arbitrary timelike trajectories, while for selected questions having control over very specific trajectories is enough. As a first illustration of  this point, let us start by recalling the definition of one of the most interesting quantities associated to a heavy probe, the cusp anomalous dimension. Consider a probe at rest that receives a sudden kick and afterwards moves with constant speed; its worldline presents a cusp at the event of the kick. The evaluation of the corresponding Wilson line will feature a logarithmic divergence \cite{polyakov}
\begin{equation}
\vev{W}\sim e^{-\Gamma_{\mbox{\tiny cusp}}(\varphi)\log\frac{L}{\epsilon}}~,
\end {equation}
with $\varphi$ the boost parameter, and $L,\epsilon$ IR and UV cutoffs, respectively. The function $\Gamma_{\mbox{\scriptsize cusp}}(\varphi)$ is the cusp anomalous dimension. This logarithmic divergence will be on top of other divergences that the vev of a similar Wilson line with a smooth contour might have. While the determination of this function for generic boosts is an interesting problem, its evaluation at either very large or very small boosts is already quite rewarding physically. Let us focus in particular in the small boost limit, by performing a Taylor expansion of the cusp anomalous dimension in  terms of the boost parameter,
\begin{equation}
\Gamma{\mbox{\scriptsize cusp}}(\varphi)=B(\lambda,N) \varphi^2+{\cal O}(\varphi^4)~.
\end{equation}
The coefficient of $\varphi^2$ was called the Bremsstrahlung function in \cite{correa}, and as indicated, it depends on the rank $N$ of the gauge group and the coupling $\lambda=g_{YM}^2 N$. At this point, we can start to illustrate the power of focusing on theories with additional symmetries.  It was argued in \cite{correa} that for probes coupled to arbitrary four-dimensional conformal field theories (CFTs), a number of physically interesting quantities are completely determined up to a coefficient, and for all these quantities, the undetermined coefficients are the same and equal to the Bremsstrahlung function. In particular, the energy loss of a probe at small velocities is
\begin{equation}
E=2\pi B\; \int dt (\dot v)^2~.
\end{equation}
This identification is valid for any line operator and any conformal field theory.

A second example of a very interesting specific trajectory is that of a probe with constant proper acceleration. Its worldline is a branch of a hyperbola, which translates into a circle in Euclidean signature. If the probe is coupled to a conformal field theory, from the vev of the circular Wilson loop and the one-point function of the stress-energy tensor in the presence of such Wilson line, one can obtain an alternative derivation of the Bremsstrahlung function \cite{fgl,lewkomalda}, the momentum difussion coefficient of the accelerated probe \cite{fgt}, or the change in entanglement entropy of a spherical region when adding a static probe \cite{lewkomalda}. In the case of  a 1/2 BPS probe of ${\cal N}=4$ $SU(N)$ super-Yang-Mills (SYM), it was argued in \cite{correa} that the Bremsstrahlung function is very simply related to the vev of a circular Wilson loop,
\begin{equation}
B=\frac{1}{2\pi^2}\lambda \partial_\lambda \hbox{log }\vev{W_{\circledcirc}}~.
\label{bfromw}
\end{equation}

Relation (\ref{bfromw}) allows us to derive the exact Bremsstrahlung function from the exact vev of a circular Wilson loop \cite{esz}, which can be evaluated by means of a matrix model computation \cite{dg} that ultimately finds its justification through localization arguments \cite{pestun}. One then finds for a fundamental 1/2 BPS probe coupled to $U(N)$ \cite{correa}
\begin{equation}
B=\frac{\lambda}{16\pi^2}\frac{L_{N-1}^2\left(-\frac{\lambda}{4N}\right)
+L_{N-2}^2\left(-\frac{\lambda}{4N}\right)}{L_{N-1}^1\left(-\frac{\lambda}{4N}\right)}~,
\label{exactb}
\end{equation}
where the $L_n^\alpha$ are generalized Laguerre polynomials. This expression was also deduced by a different chain of reasoning in \cite{fgl, lewkomalda}.

While the response of the theory to probes following these two particular trajectories is quite interesting, it is certainly far from covering all the rich physics of accelerating probes, and for other questions we need to be able to handle probes with arbitrary trajectories. In what follows we will concentrate on the study of (locally) 1/2-BPS probes of the vacuum state of ${\cal N}=4$ $SU(N)$ SYM by means of the AdS/CFT correspondence, so we find it appropriate to start by reviewing some known results.

Consider first a probe in the fundamental representation of the $SU(N)$ gauge group, i.e., a quark.  In the context of the AdS/CFT correspondence \cite{malda,gkpw}, its dual is a fundamental string embedded in AdS$_5\times \bS^5$ and governed by the Nambu-Goto action. Mikhailov \cite{mikhailov} considered such a probe following an arbitrary timelike trajectory, and found the corresponding string embedding in the supergravity regime ($N\gg 1$, $\lambda\gg 1$). The energy of the string at a given time was shown to contain both the expected instantaneous energy of the quark \cite{dragtemp}, $E_{\mbox{\scriptsize q}}=\gamma m$, and the energy that the quark has radiated over all of its previous history  \cite{mikhailov},
\begin{equation}\label{radenergyfund}
E_{\mbox{\scriptsize rad}}= \frac{\sqrt{\lambda}}{{2\pi}}\int dt_r \, a^\mu a_\mu
=\frac{\sqrt{\lambda}}{{2\pi}}\int dt_r \, \gamma^6\left(\vec a^2-|\vec v\times \vec a|^2\right)~.
\end{equation}
Notice this says that, remarkably, the total radiated power by the fundamental probe in this strongly-coupled non-Abelian theory is given (up to the coefficient) by the familiar Li\'enard formula from classical electrodynamics. The coefficient of (\ref{radenergyfund}) informs us that the Bremsstrahlung function for a 1/2-BPS particle in the fundamental representation is given in the supergravity limit by
\begin{equation}
B=\frac{\sqrt{\lambda}}{4\pi^2}~.
\end{equation}
matching the result first obtained in \cite{Kruczenski:2002fb}. In hindsight, it could have also been deduced by applying (\ref{bfromw}) to the vev of the corresponding circular Wilson loop \cite{bcfm,dgo}
\begin{equation}
\vev{W_{\circledcirc}}=e^{\sqrt{\lambda}}~.
\end{equation}
Alternatively, the same Bremsstrahlung function can be deduced from the computation of the one-point function of the Lagrangian density \cite{dkk,cg,trfsq,fgl} or energy density \cite{liusynchrotron,iancuradiation,lewkomalda,tmunu} in the presence of a quark, or the momentum diffusion coefficient of the same probe following a trajectory with constant proper acceleration \cite{xiao,brownian}.

When we move on to probes in higher-rank representations of the gauge group, the totally antisymmetric representations turn out to be the easiest generalization. In the supergravity regime, the dual of an antisymmetric $k$-quark is given \cite{yamaguchi,gp} by a D5-brane with $k$ units of worldvolume electric flux (and consequently, string charge). The brane wraps an $\bS^4\subset\bS^5$ at polar angle $\vartheta_k$ such that \cite{baryon,pr}
\begin{equation}
\sin \vartheta_k \; \cos \vartheta_k-\vartheta_k= \frac{\pi k}{N}~.
\end{equation}
Combining a very general result due to Hartnoll \cite{hartnoll} with the string solutions of Mikhailov, it is immediate to obtain the D5-brane solution dual to an antisymmetric probe following an arbitrary trajectory. This allows for the determination of the energy loss \cite{fg},
\begin{equation}
E_{\mbox{\scriptsize rad}}^{{\cal A}_k}=\frac{N\sqrt{\lambda}}{{3\pi^2}}\sin^3 \vartheta_k \; \int dt_r \,\gamma^6 \left(\vec a^2-|\vec v\times \vec a|^2\right)~.
\label{radenergyant}
\end{equation}
We observe that again the radiated power is given by a Li\'enard-type formula. The corresponding Bremsstrahlung function in the supergravity regime can be read off from (\ref{radenergyant}) to be
\begin{equation}
B_{{\cal A}_k}=\frac{N\sqrt{\lambda}}{{6\pi^3}}\sin^3 \vartheta_k~.
\end{equation}
Similarly to what happens in the fundamental representation, the exact vevs of the circular Wilson loops for probes in the antisymmetric representations have been computed exactly \cite{Fiol:2013hna}, and from them one can deduce the corresponding Bremsstrahlung function using (\ref{bfromw}).

Let us now turn to probes in the totally symmetric representation of ${\cal N}=4$ $SU(N)$ SYM, which will be the main focus of the present work.\footnote{Since the configurations we study lie at a fixed position on the $\bS^5$, our results make no use of the internal dimensions and should be relevant for other four-dimensional CFTs, whose gravity dual would involve AdS$_5$ times a different compact manifold.} In the AdS/CFT correspondence, a symmetric $k$-quark is dual to a D3-brane with $k$ units of electric flux, which is embedded fully within AdS$_5$ and reaches the boundary of this spacetime at the worldline of the probe \cite{gp,gp2}. Contrary to what happens for probes in the fundamental and antisymmetric representations, up until now the D3-brane embedding dual to an arbitrary timelike worldline was not available. This limitation was bypassed in \cite{fg} by considering a D3-brane dual to a probe with constant proper acceleration, essentially obtained in \cite{df}. The computation of the energy loss for that particular trajectory yielded the Bremsstrahlung function
\begin{equation}
B_{{\cal S}_k}=\frac{k\sqrt{\lambda}}{4\pi^2}\sqrt{1+\frac{k^2\lambda}{16\pi^2N^2}}~.
\label{bksym}
\end{equation}
Again, since the Bremsstrahlung function for this 1/2-BPS probe appears in various physical quantities, it is also possible to determine it by considering  the momentum difussion coefficient \cite{fgt} of the probe following accelerated motion, or the one-point function of the Lagrangian density in the presence of a static probe \cite{fgl}, which in turn is related by supersymmetry to the one-point function of the $\Delta=2$ chiral primary operator computed in \cite{Giombi:2006de}. The exact result for the circular Wilson loop in the totally symmetric representation was determined recently in \cite{Fiol:2013hna}, but it is given in a form that makes it very hard to carry out a systematic large $N$ expansion. The first subleading correction in $1/N$  was worked out and successfully matched to the D3-brane description very recently in \cite{leo}.

As we have stressed, there are physically interesting questions where one needs to have a handle on probes following arbitrary timelike trajectories. In the case at hand this means finding a D3-brane that reaches the boundary of AdS$_5$ at such trajectories. Besides the intrinsic interest of studying probes in higher-rank representations, there is an additional motivation that is specific to the totally symmetric case. Upon setting $k=1$ in the supergravity result (\ref{bksym}) for the $k$-symmetric Bremsstrahlung function, one obtains the result in the fundamental representation, corrected by an infinite series in $\sqrt{\lambda}/N$. What is striking is that these corrections precisely match those found in the \emph{exact} result  (\ref{exactb}) for the fundamental, in the limit where one takes $\lambda,N\rightarrow \infty$ with $\sqrt{\lambda}/N$ fixed. This was first noticed in the computation of the vev of the 1/2 BPS circular Wilson loop \cite{df}, and it is true in spite of the fact that, a priori, $k=1$ lies outside the regime of validity of the D3-brane calculation \cite{fg}. The reasons behind this better than expected behavior are not known, and in particular it is not clear what symmetries the trajectory must preserve for the D3-brane to correctly capture these subleading terms. It then seems worth exploring to what extent one can use D3-brane probe results to obtain $1/N$ corrections to other properties of probes in the fundamental representation. In order to do so, the first ingredient we need are the relevant D3-brane probes, and to those we turn next.

\subsection{Outline and main results}\label{resultssubsec}

In this paper we construct the D3-brane solution dual to a symmetric $k$-quark with arbitrary motion in the vacuum of ${\cal N}=4$ SYM. We start in Section \ref{lightstringsec} by reviewing Mikhailov's embedding (\ref{mikhsolnc}) for the string. Whereas the emphasis in many of the early applications (see, e.g., \cite{dragtemp,lorentzdirac,damping}) was on the reliance of his construction on null geodesics on the string worldsheet, which are only known after the embedding has been specified, we observe here that these curves turn out to also be null geodesics directly in spacetime (see also \cite{iancuradiation}). This implies that Mikhailov's solution can be constructed by shooting light rays in from the AdS boundary. Concretely, the string worldsheet is obtained by tracing an inbound null geodesic from each point on the quark's worldline, with tangent determined by the quark's velocity.

In Section \ref{d3subsec} we generalize this technique to the case of the D3-brane. This requires that from each point on the $k$-quark's trajectory we shoot not one but infinitely many light rays, spanning the surrounding $\bS^2$. This ultimately leads to the 4-dimensional embedding (\ref{D3covariant}), which, when combined with a worldvolume field strength chosen to be proportional to the induced metric, as specified in (\ref{Fg}), quite remarkably turns out to satisfy all of the (highly nonlinear) equations of motion. We thus achieve our main goal in the paper, to obtain the heretofore unknown embeddings dual to $k$-symmetric probes with an arbitrary timelike worldline. We illustrate the general construction with the explicit example of a probe in circular motion.

The eminently geometric character of our construction method suggests that it should have wider relevance, and indeed, in Section \ref{d5subsec} we show that the known D5-brane embeddings dual to $k$-antisymmetric probes are correctly reproduced by tracing light rays in from the AdS boundary and selecting the field strength according to (\ref{Fg}). Many other extensions seem possible and desirable, and we intend to pursue them in future work.

As a first application of our D3-brane solutions, we compute their total energy and show that, much as in \cite{mikhailov,dragtemp}, it cleanly splits into two contributions attributable respectively to the $k$-quark's intrinsic and radiated energy. For the former, we find
\begin{equation}
E_{k\mbox{\scriptsize q}}=k\gamma m~,
\label{intenergysym}
\end{equation}
meaning that even for arbitrary motion our probe continues to be a threshold bound state. For the latter, we obtain
\begin{equation}
E_{\mbox{\scriptsize rad}}^{{\cal S}_k}=\frac{k\sqrt{\lambda}}{2\pi}\sqrt{1+\frac{k^2\lambda}{16\pi^2N^2}} \int dt_r
\left[\gamma^6\left(\vec a^2-|\vec v\times \vec a|^2\right) \right]~.
\label{radenergysym}
\end{equation}
which has the by now familiar Li\'enard-type form, and confirms the Bremsstrahlung function (\ref{bksym}) deduced previously for a probe with proper constant acceleration. This shows in particular that, at least for this observable, the relation between the fundamental and totally symmetric representations extends to arbitrary probe trajectories.

\section{Strings of Light}\label{lightstringsec}

 We focus on the duality between maximally supersymmetric Yang-Mills theory (MSYM) in $3+1$ dimensions, with gauge group $SU(N)$, and Type IIB string theory on 5-dimensional anti-de Sitter (AdS) spacetime cross a 5-sphere. To examine the gauge theory on Minkowski spacetime, we work in the Poincar\'e patch of AdS$_5$, where the metric reads
\begin{equation}\label{metric}
G_{mn}dx^m dx^n=\frac{L^2}{z^2}\left(\eta_{\mu\nu}dx^{\mu}dx^{\nu}+dz^2\right)
=\frac{L^2}{z^2}\left(-dt^2+dx_1^2+dx_2^2+dx_3^2+dz^2\right)~.
%\left(-dt^2+dr^2+r^2 d\theta^2+r^2\sin^2 \theta d\phi^2+dz^2\right)\,,
\end{equation}
 There are also a constant dilaton $e^{\phi}=g_s$ and $N$ units of flux of the self-dual Ramond-Ramond 5-form field strength. The AdS component of the corresponding
potential can be taken to be
\begin{equation}\label{C4}
C_{0123}=-\frac{L^4}{z^4}~.
\end{equation}
The ratio of the AdS radius of curvature $L$ to the string length $\ls$ is related to the MSYM 't~Hooft coupling through
\begin{equation}\label{L}
\frac{L}{\ls}=\lambda^{1/4}~.
\end{equation}

 An external (infinitely massive) quark in MSYM is dual to a fundamental string extending from the AdS boundary at $z=0$ to the Poincar\'e horizon at $z\to\infty$. More precisely, it is the endpoint of the string that is dual to the quark, whereas the body of the string encodes the profile of the gluonic (and other) field(s) sourced by the quark. Given a quark trajectory $x^{\mu}(\tau)$, the dual string embedding $X^{\mu}(\tau,z)$ is found by extremizing the usual Nambu-Goto action, subject to the condition that the string endpoint follow the same path as the quark, $X^{\mu}(\tau,0)=x^{\mu}(\tau)$. Remarkably, Mikhailov \cite{mikhailov} was able to solve this problem for \emph{arbitrary} quark motion. His solution takes the surprisingly simple form
 \begin{equation}\label{mikhsol}
 X^{\mu}(\tau,z)=x^{\mu}(\tau)+v^{\mu}(\tau)z~,
 \end{equation}
 where $\tau$ is chosen to be the proper time of the quark, and $v^{\mu}\equiv dx^{\mu}/d\tau$ denotes its 4-velocity. This solution is \emph{retarded} or \emph{purely outgoing}, i.e., it is appropriate for the situation of primary physical interest, where excitations of the MSYM fields propagate outward from the quark toward infinity. Flipping the sign in front of $v^{\mu}$ gives instead an advanced or purely ingoing solution.\footnote{The more generic embedding would be a nonlinear superposition of retarded and advanced contributions, but its form is not known.}

 The physics of the embedding (\ref{mikhsol}) is more easily unpacked by rewriting it in noncovariant notation,
 \begin{eqnarray}\label{mikhsolnc}
 t(t_r,z)&=&t_r+\gamma(t_r)z~,
 \\
 \vec{X}(t_r,z)&=&\vec{x}(t_r)+\gamma(t_r)\vec{v}(t_r)z~.
 \nonumber
 \end{eqnarray}
 The information of interest here is the position at time $t$ of the string bit at radial depth $z$, which essentially codifies the configuration at that time of the gluonic field a distance $\sim z$ away from the quark. We see from the second equation that this information is determined by the position $\vec{x}$ and 3-velocity $\vec{v}$ of the quark/endpoint at the earlier, \emph{retarded} time $t_r$ defined by the first equation, in close analogy with the Li\'enard-Wiechert story in classical electromagnetism.\footnote{The fact that, unlike in classical electromagnetism, the gluonic field is a nonlinear medium that can rescatter signals is accounted for in the dual gravity description by the fact that fields at the AdS boundary receive contributions from each and every point along the string \cite{cg,trfsq}. Even so, 1-point functions of gluonic field observables turn out to have some surprising features, such as beaming \cite{liusynchrotron,veronika,beaming}, lack of radial/temporal broadening \cite{iancuaspects,iancuradiation,trfsq} and a `near'-field tail reaching out to infinity \cite{tmunu,lewkomalda}.}

 The curves at constant $t_r$ (or $\tau$) are null geodesics on the string worldsheet, and this property played an important role in Mikhailov's construction. But what will be of interest to us in this paper is the observation that they are also null geodesics in spacetime, as is clear from (\ref{mikhsolnc}), where we see that they are straight lines in the Poincar\'e coordinates (\ref{metric}). We can therefore interpret the Mikhailov embedding as the surface obtained by shooting light rays from the AdS boundary into the bulk, with the starting point and slope of each ray determined respectively by the position and velocity of the quark, according to (\ref{mikhsol}) or (\ref{mikhsolnc}). The nontrivial fact that the ruled surface so obtained is a classical string worldsheet (i.e., a solution of the Nambu-Goto system) was proven somewhat indirectly in \cite{mikhailov}, by arguing that it extremizes the action, and later verified in \cite{dampingtemp} by directly plugging it into the equations of motion.

A connection between null geodesics and strings was discussed previously in studies of light parton energy loss in a thermal gluon plasma \cite{gubserlight,jensenlight1,jensenlight2,arnold}. Light rays were employed there to estimate the trajectory of the endpoint of the string dual to a light quark or gluon, as it falls towards the black hole horizon (see however \cite{horowitz}). In contrast, we are here finding that (at least in pure AdS) the entire worldsheet of the string dual to a heavy quark can be generated by throwing light rays in from the boundary.\footnote{Another interesting use of light rays in the holographic description of quarks appeared in the beaming proposal of \cite{veronika} to determine the gravitational backreaction of the string (see also \cite{beaming}). A relation between light rays and the string worldsheet was also postulated in \cite{zahed} in the finite temperature context, but it was not shown whether the ansatz in question satisfies the equation of motion.} It is perhaps worth emphasizing that the motion of internal points of the string, unlike that of its endpoint, depends on the choice of worldsheet coordinates. To uncover the geometric description of the string in terms of light rays, it is crucial then that in (\ref{mikhsolnc}) (or (\ref{mikhsol}))  we are choosing to parametrize the worldsheet in terms of the (proper) retarded time $t_r$ ($\tau$) at the boundary.

 \section{Branes of Light}\label{lightbranesec}

Having understood the purely geometric origin of the Mikhailov solution for the string, it is natural to wonder whether it might be possible to construct other brane embeddings by similarly shooting light from the AdS boundary. In this section we will examine this question for the specific case of a D3-brane with $k$ units of electric flux, which is known to be dual to a $k$-quark in the totally symmetric representation of the $SU(N)$ group \cite{df,gp,gp2,dgrt}. As explained in the Introduction, this case is of particular interest because up to  now the appropriate D3-brane embeddings are only known for a couple of specific $k$-quark worldlines, in contrast with the case of the totally antisymmetric representation. The latter is dual \cite{baryon,yamaguchi,gp,draggluon} to a D5-brane wrapped on a $\bS^4\subset \bS^5$ \cite{pr,cpr}, whose embedding in AdS turns out to exactly coincide \cite{hartnoll}, for arbitrary $k$-quark trajectory, with the string solution (\ref{mikhsol}) dual to a quark with the same trajectory. We will come back to that case at the end of the section.

\subsection{D3-brane}\label{d3subsec}

Let us begin by considering the case of a static $k$-quark, $\vec{x}(t)=\mbox{constant}$. Our goal is to generate the 4-dimensional worldvolume of the dual D3-brane by shooting light rays into the bulk as we move along this worldline at the AdS boundary. In Poincar\'e coordinates, each light ray is a straight line, and will correspond to a fixed retarded time $t_r$. By symmetry, it is clear that identical light rays must be shot along all directions $(\theta,\phi)$ of the $\bS^2$ that surrounds the location of the $k$-quark. We thus obtain an ansatz for the D3 embedding that is closely analogous to (\ref{mikhsolnc}):
\begin{eqnarray}\label{D3static}
\vec{X}(t_r,z,\theta,\phi)&=&\vec{x}(t_r)+\kappa z\vec{n}~,
\\
t(t_r,z,\theta,\phi)&=&t_r+\sqrt{1+\kappa^2}\,z~,
\nonumber
\end{eqnarray}
where $\vec{n}$ for now denotes the unit vector $(\cos\theta,\sin\theta\cos\phi,\sin\theta\sin\phi)$, $\kappa$ is a number expressing the common slope of all of the straight lines, and the $\kappa$-dependence of the last equation has been fixed with the requirement that these lines be null. To be able to represent a bound state of $k$ quarks, we expect this putative D3-brane to carry $k$ units of fundamental string charge along the AdS radial direction. In other words, the embedding should be sustained by $k$ units of the worldvolume electric flux  associated with $F_{t_r z}$ (this will be made more explicit in (\ref{Pitz}) below, after we write down the D3-brane action). The slope $\kappa$ must depend on $k$, because for $k=1$ (and $N\to\infty$) all lines must come together as the D3-brane embedding closes up into the fundamental string (\ref{mikhsolnc}) known to be dual to a static quark.

The D3-brane solution for the static case was obtained in \cite{reyyee, df}, and indeed happens to agree with our ansatz (\ref{D3static}), with
 \begin{equation}\label{kappa}
\kappa=\frac{k\sqrt{\lambda}}{4N}~, \nonumber
\end{equation}
and
\begin{equation}\label{Ftzstatic}
F_{t_r z}=\frac{\sqrt{\lambda}}{2\pi}\frac{1}{z^2}~.
\end{equation}
This serves then as a first successful test of our approach.

\begin{figure}[hbt]
\begin{center}
  \includegraphics[width=8cm]{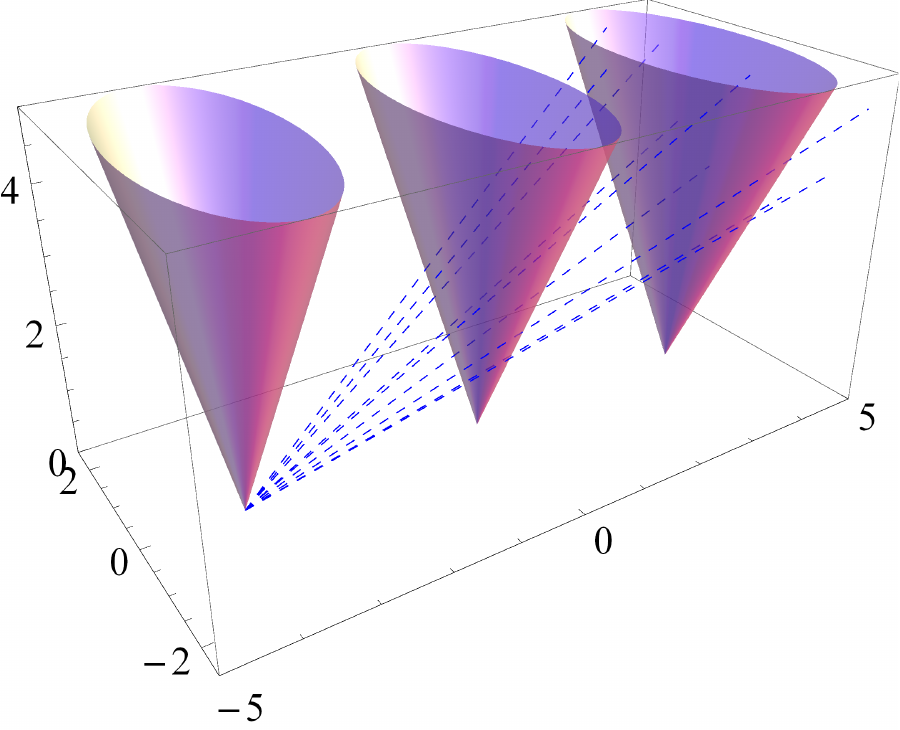}
  \setlength{\unitlength}{1cm}
\begin{picture}(0,0)
\put(-4.3,0.9){ $x_1$}
\put(-5.1,0.5){\vector(2,1){0.8}}
\put(-8.1,1.7){ $x_2$}
\put(-7.3,0.7){\vector(-1,2){0.4}}
\put(-8.9,5.2){ $z$}
\put(-8.4,4.2){\vector(-1,4){0.2}}
\end{picture}
\end{center}
\vspace*{-0.8cm}
\caption{Successive snapshots (at $t=-4,0,4$) of the D3-brane solution dual to a $k$-quark moving at constant velocity along the $x^1$ direction (color online), with $v=0.9$ and $\kappa=1/2$. The embedding is symmetric under rotations in the $x^2$-$x^3$ plane, and the boosted conical surfaces shown here are just the  sections at fixed azimuthal angle $\phi=0$. The dashed blue lines are light rays emitted into AdS from the boundary at $t=-4$. The entire worldvolume is generated by such ingoing rays, emitted from all points along the trajectory of the $k$-quark.
\label{vfig}}
\end{figure}
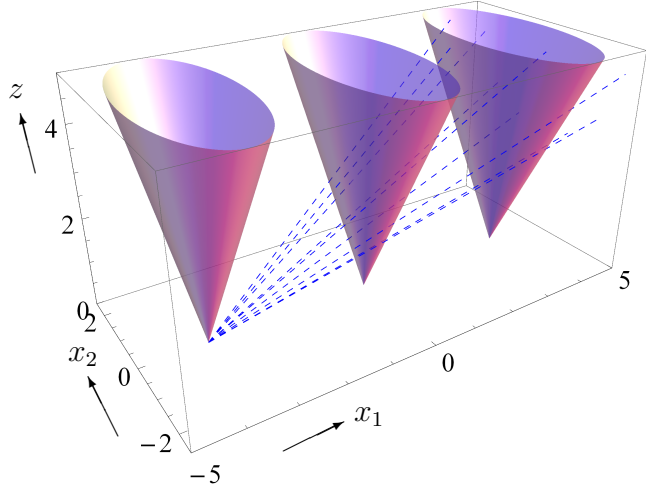

Notice that (\ref{D3static}) can be rewritten as
\begin{equation}\label{D3constantv}
X^{\mu}(\tau,z,\theta,\phi)=x^{\mu}(\tau)+\left[\sqrt{1+\kappa^2}\,v^{\mu}+\kappa n^{\mu}(\theta,\phi)\right]z~,
\end{equation}
with $v^{\mu}=(1,0,0,0)$ the 4-velocity of the static $k$-quark and $n^{\mu}\equiv(0,\vec{n})$. By Lorentz covariance, we are assured then that this same equation gives the correct D3-brane embedding dual to a $k$-quark translating uniformly, as long as we take $v^{\mu}$ to be the corresponding velocity and $n^{\mu}$ the appropriately boosted 4-vector. Note that the latter automatically satisfies the two constraints $n^2=1$ and $n\cdot v=0$, and so has two independent components, which we are choosing to parametrize with the original rest frame angles $(\theta,\phi)$.  The resulting embedding is shown at various times in Fig.~\ref{vfig}, along with some of the null geodesics that generate the D3-brane worldvolume. The corresponding  field strength is
\begin{equation}\label{Ftauz}
F_{\tau z}=\frac{\sqrt{\lambda}}{2\pi}\frac{1}{z^2}~,
\end{equation}
i.e., $F_{t_r z}=\sqrt{\lambda}/2\pi\gamma z^2$. Thanks to our choice of worldvolume coordinates, no magnetic components are turned on in spite of the boost.

In Mikhailov's solution (\ref{mikhsol}), the string embedding depends only on the quark's position and velocity. It is natural to suspect that the same is true for our case, and we are led then to conjecture that allowing for a time-dependent velocity in (\ref{D3constantv}),
\begin{equation}\label{D3covariant}
X^{\mu}(\tau,z,\theta,\phi)=x^{\mu}(\tau)+\left[\sqrt{1+\kappa^2}\,v^{\mu}(\tau)+\kappa n^{\mu}(\tau,\theta,\phi)\right]z~,
\end{equation}
yields the correct D3-brane embedding dual to a $k$-quark with an \emph{arbitrary} trajectory. The $\tau$-dependence in $n^{\mu}$ is due to the fact that this 4-vector has to be suitably transported along the worldline. To avoid confusion, we will henceforth denote the rest frame unit 3-vector by $\vec{n}_R$, and let $\vec{n}$ stand for the spatial part of the Lorentz-transformed 4-vector $n^{\mu}$. Notice that, due to the properties of this 4-vector stated in the previous paragraph, the proposed embedding (\ref{D3covariant}) satisfies $(X-x)^2=-z^2$, which is a restatement of the fact that light fronts are being emitted from $z=0$ at each $x^{\mu}(\tau)$.  This equation is also satisfied by Mikhailov's string embedding, (\ref{mikhsol}).

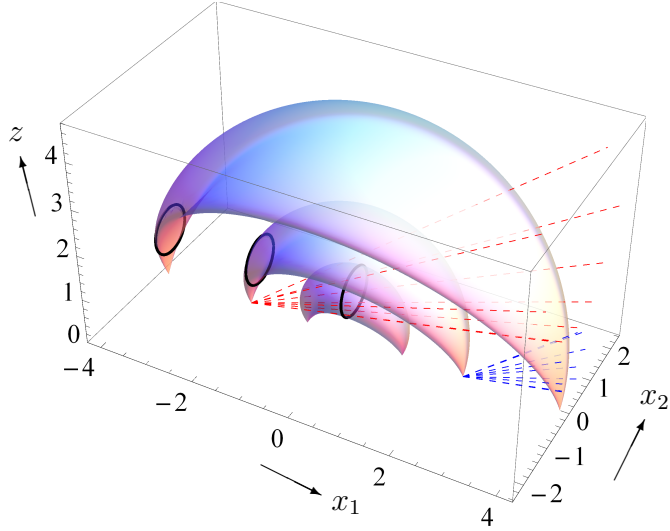
\begin{figure}[ht]
\begin{center}
  \includegraphics[width=8cm]{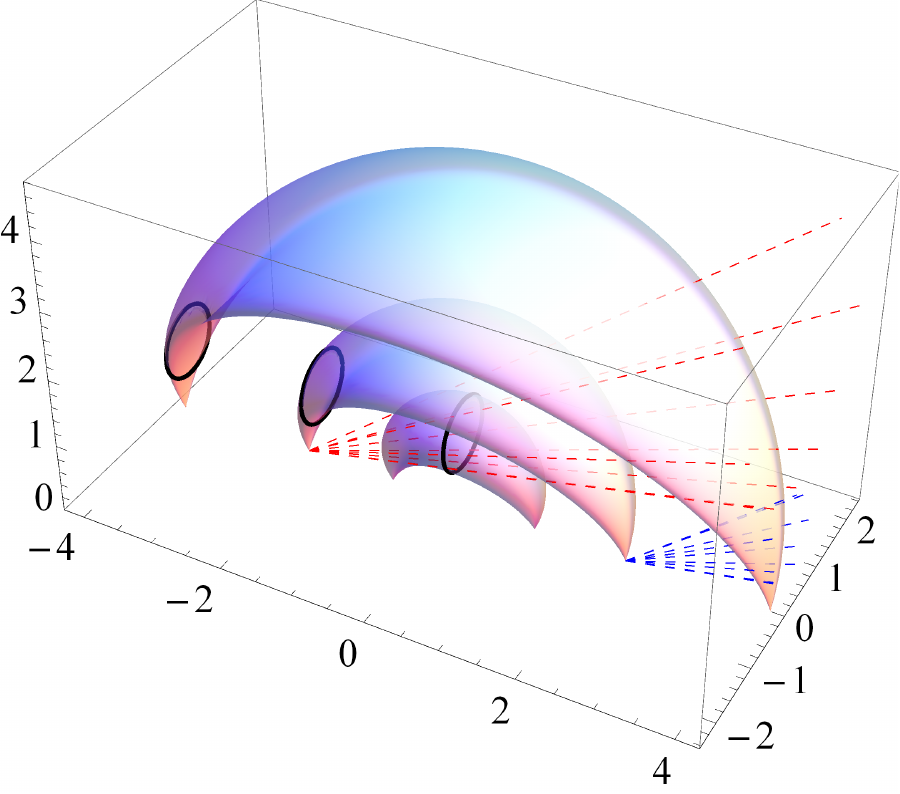}
    \setlength{\unitlength}{1cm}
\begin{picture}(0,0)
\put(-4.6,0.3){ $x_1$}
\put(-5.4,0.9){\vector(2,-1){0.8}}
\put(-0.5,1.7){ $x_2$}
\put(-0.7,0.7){\vector(1,2){0.4}}
\put(-8.9,5.2){ $z$}
\put(-8.4,4.2){\vector(-1,4){0.2}}
\end{picture}
\end{center}
\vspace*{-0.8cm}
\caption{Successive snapshots (at $t=0,2,4$) of the D3-brane solution dual to a $k$-quark and $k$-antiquark undergoing back-to-back uniform acceleration along $x^1$, with $A=1$ and $\kappa=1/2$. The embedding is invariant under azimuthal rotations, and the banana-shaped surfaces shown here are just the $\phi=0$ sections. The dashed blue (red) lines are light rays emitted into AdS from (absorbed from AdS into) the $k$-quark ($k$-antiquark) tip of the brane, at $t=2$. The full worldvolume is generated by all such ingoing (outgoing) rays.
The black circle marks the locus where the retarded quark and advanced antiquark embeddings smoothly join together, which is seen to move at the speed of light in the negative $x^1$ direction (see Appendix~\ref{accelapp}).
\label{afig}}
\end{figure}

To visualize our ansatz more explicitly, we can spell it out when the  motion is purely
along direction $x\equiv x^1$:
\begin{eqnarray}\label{D31D}
t&=&t_r+\gamma z(\sqrt{1+\kappa^2}+\kappa v\cos\theta )~,
\nonumber\\
X^1&=&x+\gamma z(\sqrt{1+\kappa^2}v+\kappa\cos\theta)~,
\nonumber\\
X^2&=&\kappa z\sin\theta\cos\phi~,
\\
X^3&=&\kappa z\sin\theta\sin\phi~.
\nonumber
\end{eqnarray}
For the specific case of uniform proper acceleration $A$, $x(t)=\sqrt{A^{-2}+t^2}$,
%(which is Wick-related to a Euclidean circle),
this can be compared against the solution obtained in \cite{df,fg}, which was found to be described by
\begin{equation}\label{D3accel}
\left(-t^2+\vec{X}^2+z^2-A^{-2}\right)^2+4 A^{-2}\left(X_2^2+X_3^2\right)-4\kappa^2 A^{-2}z^2=0~.
\end{equation}
And indeed, the ansatz (\ref{D31D}) can be verified to satisfy this equation, which constitutes a second successful test of our construction. The embedding at various times is shown in Fig.~\ref{afig}, together with some of the corresponding light rays. In parallel with the quark/string case examined in \cite{noline,eprer}, the retarded D3-brane solution (\ref{D31D}) associated with a uniformly accelerated $k$-quark terminates at a finite radial position, and can only be smoothly completed by a suitable \emph{advanced} $k$-antiquark solution. More details on this are given in  Appendix~\ref{accelapp}.

The worldvolume $U(1)$ gauge field should still encode $k$ units of the electric flux associated with $F_{t_r z}$. But for this we get no guidance from Mikhailov's solution, so it is \emph{a priori} unclear whether (\ref{Ftauz}) continues to be the only nonvanishing component of the field strength, or if other components are turned on, due to some acceleration-dependent contributions.  To settle this point, we must work out and enforce the relevant equations of motion.

Let us turn then to the D3-brane action, which includes both the Dirac-Born-Infeld and Wess-Zumino terms,
\begin{equation}\label{D3action}
S_{\text{D3}}=T_{\text{D3}}\int d^4\xi\, \left(-\sqrt{-\det(g_{\alpha\beta}+2\pi \ls^2 F_{\alpha\beta})}+c_{0123}\right)~,
\end{equation}
where
$T_{\text{D3}}=1/(2\pi)^3\gs\ls^4=N/2\pi^2 L^4$ is the tension of the D3-brane,
$g_{\alpha\beta}=\p_{\alpha} X^m \p_{\beta} X^{n}G_{mn}$ is the induced metric,
and $c$ is likewise the pullback of the 4-form $C$ onto the worldvolume,
$c_{\alpha\beta\gamma\delta}=\p_{\alpha} X^m\p_{\beta} X^n\p_{\gamma} X^p\p_{\delta} X^q C_{mnpq}$.
The worldvolume coordinates are denoted collectively by $\xi^{\alpha}$, and, based on the preceding analysis, will be chosen by us to be
 $t_r$, $z$, $\theta$, and $\phi$.

 {}From this action we can work out the momentum density conjugate to each of the embedding fields,
 $P^{\alpha}_m\equiv \p\cL_{\text{D3}}/\p(\p_{\alpha} X^m)$, and to each of the gauge field components, $\Pi^{\alpha\beta}\equiv\p\cL_{\text{D3}}/\p(\p_{\alpha} A_{\beta})=\p\cL_{\text{D3}}/\p F_{\alpha\beta}$. The generic expressions are long and unenlightening, so we will not write them here. Since the fields $X^{\mu}$ and $A_{\alpha}$ do not appear undifferentiated in $S_{\text{D3}}$, their equations of motion are just the statement that these momenta are conserved:
 \begin{equation}\label{eom}
 \p_{\alpha} P^{\alpha}_{\mu}=0~,\qquad\p_{\alpha}\Pi^{\alpha\beta}=0~.
 \end{equation}

 Consider first the case of motion purely along one dimension. Using our ansatz (\ref{D31D}) for the embedding, and assuming for the time being that (\ref{Ftauz}) is the only nonvanishing field strength component, we obtain
 \begin{equation}\label{Pitz}
 \Pi^{t_r z}=\frac{k\sin\theta}{4\pi}~.
 \end{equation}
 Upon angular integration, this gives the desired integer value for the fundamental string charge (along the radial AdS direction) carried by the D3, $\int d\theta d\phi\,\Pi^{t_r z}=k$, thereby validating (\ref{D31D}) and (\ref{Ftauz}). In fact, using (\ref{D31D}) one finds that in order for (\ref{Pitz}) to hold, all magnetic components of the field strength must be set to zero, $F_{z\theta}=F_{z\phi}=F_{\theta\phi}=0$, but the electric components $F_{t_r \theta}$ and $F_{t_r \phi}$ can be arbitrary. Independently of the values of the latter, we find that $\Pi^{t_r \theta}$ and $\Pi^{t_r \phi}$  vanish, just as they should.

 Allowing for these electric components of $F$ to be arbitrary, and continuing to use (\ref{D31D}) and (\ref{Ftauz}), we can then notice that
 $\Pi^{\theta\phi}=0$ automatically, but
 \begin{eqnarray}\label{Pismagnetic}
 %kappa=k*sqrt{lambda}/4N, so k=4N*kappa/sqrt{lambda}
 \Pi^{z\theta}&=&\frac{4N\sqrt{1+\kappa^2}\gamma^2\sin\theta}{4\pi \lambda\kappa}
 \left[\frac{2\pi}{\gamma^2}F_{t_r\theta}-\sqrt{\lambda}\kappa a\sin\theta \right]~,
 \nonumber\\
 \Pi^{z\phi}&=&\frac{2N\sqrt{1+\kappa^2}\csc\theta}{\lambda\kappa}F_{t_r\phi}~,
 \end{eqnarray}
 with $a$ the acceleration of the $k$-quark.
 These momenta must be set to zero, because they would represent unwanted D1-brane charge densities on the D3, and more importantly, because if they do not vanish then the gauge field equations of motion (\ref{eom}) are not satisfied. Notice that this forces us to turn on $F_{t_r\theta}$, with a value proportional to the acceleration, which is consistent with the fact that this component was not found to be excited when the $k$-quark moves with constant velocity.

Altogether, we have deduced then that, in the case of one-dimensional motion, the worldvolume field strength must take the form
\begin{eqnarray}\label{F1D}
F_{t_r z}&=&\frac{\sqrt{\lambda}}{2\pi}\frac{1}{\gamma z^2}~,
\nonumber\\
F_{t_r \theta}&=&\frac{\sqrt{\lambda}}{2\pi}\kappa\gamma^2 a\sin\theta~,
\\
F_{t_r\phi}&=&0~,
\nonumber\\
F_{z\theta}&=&F_{z\phi}=F_{\theta\phi}=0~.
\nonumber
\end{eqnarray}
And as a definitive, highly nontrivial test of our overall ansatz, one can verify that all of the equations of motion (\ref{eom}) are correctly satisfied upon assuming (\ref{D31D}) and (\ref{F1D}).

It will prove useful to note that the field strength (\ref{F1D}) is closely related to the induced metric on the D3-brane,
\begin{equation}\label{Fg}
2\pi \ls^2 F_{\alpha\beta}=-\frac{g_{\alpha\beta}}{\sqrt{1+\kappa^2}}\qquad\forall\quad \alpha<\beta~.
\end{equation}
This relation, the fact that $g_{zz}=0$ (which is nothing but the requirement that each line at constant $t_r,\theta,\phi$ be null), and the vanishing  of all off-diagonal space-space components of the induced metric are crucial features of the solution. Together, they imply the drastic simplifications
\begin{equation}\label{bisimp}
\sqrt{-\det(g_{\alpha\beta}+2\pi \ls^2 F_{\alpha\beta})}=\frac{\kappa}{\sqrt{1+\kappa^2}}\sqrt{-\det g_{\alpha\beta}}
\end{equation}
and
\begin{equation}\label{ngsimp}
\sqrt{-\det g_{\alpha\beta}}=-g_{t_r z}L^2 \kappa^2 \mbox{vol}(\bS^2)~,
\end{equation}
where $\mbox{vol}(\bS^2)=\sin\theta$.

Let us now move on to the case of arbitrary 3-dimensional motion. To give meaning to (\ref{D3covariant}), we need a precise specification of the unit vector $n^{\mu}(\tau,\theta,\phi)$, or in other words, a choice of labels $(\theta,\phi)$ for the light rays shot along the different directions of the $\bS^2$ that surrounds $\vec{x}(\tau)$.
%coordinates $(\theta,\phi)$ on the $\bS^2$ that surrounds $\vec{x}(\tau)$.
The timelike vector at rest $v_R^{\mu}=(1,0,0,0)$ is converted into an arbitrary 4-velocity $v^{\mu}$  by the Lorentz transformation
 \begin{equation}\label{boost}
\Lambda^{\mu}{}_{\nu}=
\left(
\begin{array}{cccc}
\gamma & \gamma v_1 & \gamma v_2 & \gamma v_3 \\
\gamma v_1 & 1+\frac{\gamma^2 v_1^2}{1+\gamma} & \frac{\gamma^2 v_1 v_2}{1+\gamma} & \frac{\gamma^2 v_1 v_3}{1+\gamma} \\
\gamma v_2 & \frac{\gamma^2 v_1 v_2}{1+\gamma} & 1+\frac{\gamma^2 v_2^2}{1+\gamma} & \frac{\gamma^2 v_2 v_3}{1+\gamma} \\
\gamma v_3 & \frac{\gamma^2 v_1 v_3}{1+\gamma} & \frac{\gamma^2 v_2 v_3}{1+\gamma} & 1+\frac{\gamma^2 v_3^2}{1+\gamma}
\end{array} \right)~,
 \end{equation}
 where $v_i$ denotes the components of the 3-velocity. A subtlety in the case of accelerated motion along more than one dimension, however, is that if we use this  canonical boost based on $v^{\mu}(\tau)$ to define $n^{\mu}(\tau,\theta,\phi)$ at each point along the worldline of the $k$-quark, the outcomes at successive points would not be related purely by a boost, but would also include a rotation (this fact is at the root of the Thomas precession). The most natural choice, then, is to avoid this spurious rotation on the $\bS^2$, by demanding that $n(\tau,\theta,\phi)$ be transported along the worldline as dictated by the Fermi-Walker equation (see, e.g., \cite{misner})
 \begin{equation}\label{fw}
\p_{\tau}n^{\mu}=(n\cdot\p_{\tau}v) v^{\mu}-(n\cdot v)\p_{\tau}v^{\mu}~,
\end{equation}
from which the second term drops out because $n\cdot v=0$. In more detail, at some initial time $\tau_0$ we apply the boost (\ref{boost}) to define $n^{\mu}(\tau_0,\theta,\phi)=\Lambda^{\mu}{}_{\nu}n^{\mu}_R(\theta,\phi)$, and then evolve from there using (\ref{fw}) to obtain $n^{\mu}$ at arbitrary times. Note that the result generally depends on the probe's velocity \emph{and} acceleration.

For use below, it is useful to notice that for each $\tau$, the 4 vectors $v^{\mu}$, $n_1^{\mu}\equiv n^{\mu}$, $n_2^{\mu}\equiv \p_{\theta}n^{\mu}$ and $n_3^{\mu}\equiv\p_{\phi}n^{\mu}/\sin\theta$ form an orthonormal tetrad,
\begin{equation}\label{tetrad}
v^2=-1~,\qquad v\cdot n_I=0~,\qquad  n_I\cdot n_J =\delta_{IJ}~,
\end{equation}
with each $n_I$ transported according to (\ref{fw}), and with the additional property
\begin{equation}\label{tetradlevi}
\epsilon_{\mu \nu \lambda \rho} \; v^\mu n_1^\nu n_2^\lambda n_3 ^\rho=1~.
\end{equation}
Physically, the implication is that these four vectors define a nonrotating reference frame.
It is easy to check that (\ref{tetrad}) and (\ref{tetradlevi}) hold in the rest frame, and they then follow in general from the fact that the angular derivatives commute both with the boost (\ref{boost}) and with the transport (\ref{fw}). Finally, it follows from the definition (\ref{fw}) of Fermi-Walker transport that
\begin{equation}
\p_{t_r}n_I\cdot n_J =0= \p_{t_r}v\cdot\p_{t_r}n_I~,
\end{equation}
and we will make repeated use of these relations in what follows.

 The calculation of the induced metric on the embedding (\ref{D3covariant}) is presented in Appendix~\ref{inducedapp}.
 Upon examining the momentum densities conjugate to the gauge field, the situation is found to be exactly the same as in the 1-dimensional case: all magnetic components of the worldvolume field strength must be set to zero in order to reproduce the expected $k$ units of string charge (\ref{Pitz}), and all other components of $\Pi^{\alpha\beta}$ vanish automatically except for $\Pi^{z\theta}$ and $\Pi^{z\phi}$. Setting these to zero we thus obtain again two equations for the two unknown electric components of $F$. Just as for 1-dimensional motion, the solution takes the form (\ref{Fg}). More explicitly, we must set
\begin{eqnarray}\label{F3D}
F_{t_r z}&=&\frac{\sqrt{\lambda}}{2\pi}\frac{1}{\gamma z^2}~,
\nonumber\\
F_{t_r \theta}&=&-\frac{\sqrt{\lambda}}{2\pi}\kappa
\p_{t_r}v\cdot n_2~,
\\
F_{t_r\phi}&=&-\frac{\sqrt{\lambda}}{2\pi}\kappa\sin\theta
\p_{t_r}v\cdot n_3~,
\nonumber\\
F_{z\theta}&=&F_{z\phi}=F_{\theta\phi}=0~.
\nonumber
\end{eqnarray}

Again it is true that the $zz$ and  off-diagonal space-space components of the induced metric vanish, so (\ref{bisimp}) and (\ref{ngsimp}) still hold (see Appendix~\ref{inducedapp}).
And again, the ultimate test is the verification that with (\ref{D3covariant}) and (\ref{Fg}) (or (\ref{F3D})) all equations of motion are correctly satisfied.
We have thus succeeded in constructing our desired D3-brane embeddings by shooting light in from the AdS boundary.

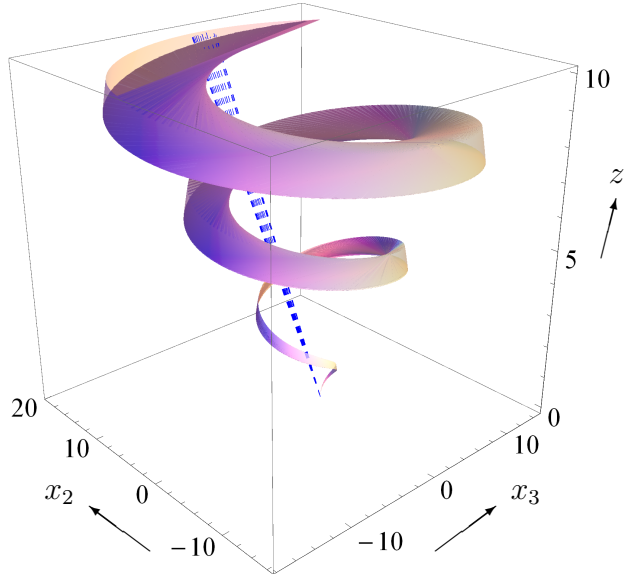
\begin{figure}[ht]
\begin{center}
  \includegraphics[width=8cm]{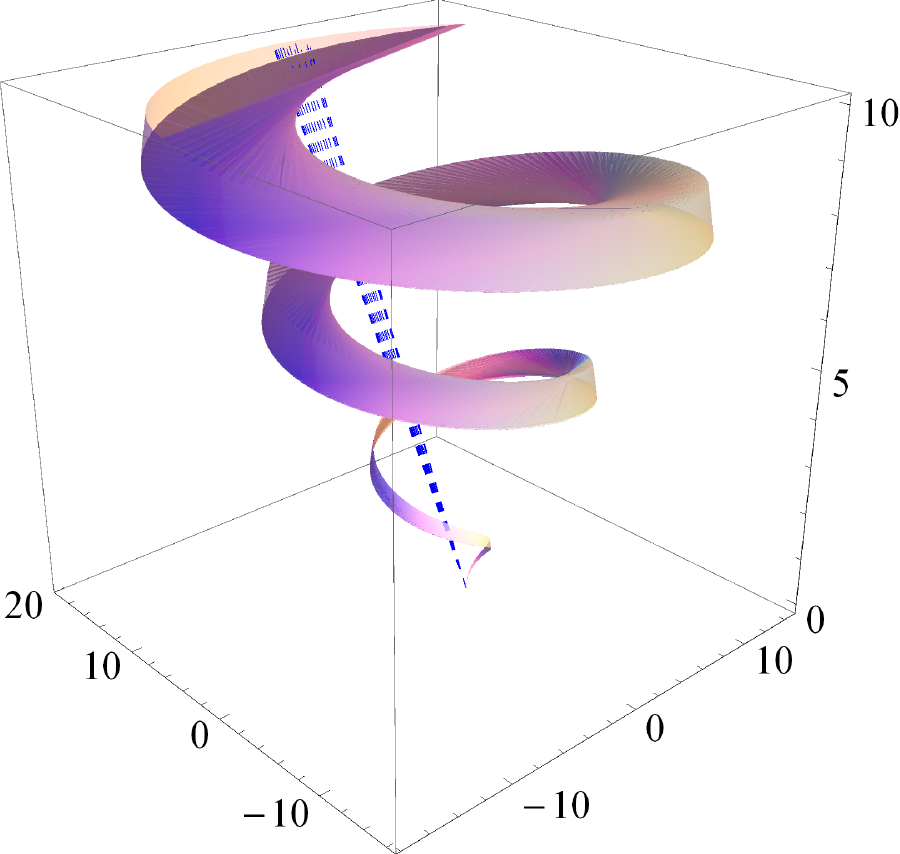}
    \setlength{\unitlength}{1cm}
\begin{picture}(0,0)
\put(-7.8,1.0){$x_2$}
\put(-6.4,0.3){\vector(-4,3){0.8}}
\put(-1.6,1.0){$x_3$}
\put(-2.6,0.3){\vector(4,3){0.8}}
\put(-0.3,5.2){$z$}
\put(-0.4,4.2){\vector(1,4){0.2}}
\end{picture}
\end{center}
\vspace*{-0.8cm}
\caption{Snapshot at $t=0$ of the D3-brane solution dual to a $k$-quark undergoing uniform circular motion on the $x^2$-$x^3$ plane, with radius $R=1$, angular frequency $\omega=7/8$ and $\kappa=0.1$. The plot omits $x^1$ and shows only the section of the embedding at polar angle $\theta=\pi/3$, over the full range of the azimuthal angle $\phi$. The dashed blue lines are light rays emitted into AdS from the $k$-quark at $t=0$. The full worldvolume is generated by all such ingoing rays.
\label{circularfig}}
\end{figure}

As an example, consider the case of motion with constant angular velocity $\omega$ on a circle of radius $R$ on the $x^2$-$x^3$ plane: $v^{\mu}=\gamma(1,0,-\omega R\sin\omega t,\omega R\cos\omega t)$. Solving the Fermi-Walker equation (\ref{fw}) one finds \cite{misner}
\begin{eqnarray}\label{fwcircular23}
n^0&=&\gamma\omega R \sin \theta \sin (\phi-\gamma \omega t)~,
\nonumber\\
n^1&=& \cos \theta~,
\\
n^2&=&\sin \theta\left[\cos \omega t \cos (\phi-\gamma \omega t)-\gamma \sin \omega t \sin(\phi-\gamma \omega t)\right]~,
\nonumber\\
n^3&=&\sin\theta \left[\sin \omega t \cos (\phi-\gamma \omega t)+\gamma \cos \omega t \sin(\phi-\gamma \omega t)\right]~.
\nonumber
\end{eqnarray}
The resulting embedding (\ref{D3covariant}) is depicted in Fig.~\ref{circularfig}.

\subsection{D5-brane}\label{d5subsec}

Encouraged by our success with the D3-brane, it is natural to suspect that the essential idea of the method, tracing out the embedding of interest using null spacetime geodesics, should be equally useful to construct branes of various types in diverse dimensions, probably even on backgrounds other than pure AdS. As a first step in this direction, we now show that the D5-brane solutions dual to a $k$-quark in the totally antisymmetric representation of $SU(N)$ do in fact fit naturally within the same framework.

For the static case, the embeddings we have in mind were deduced in \cite{pr,cpr}, and then understood in \cite{draggluon} to be a special case of the larger family of solutions constructed earlier in \cite{baryon}. Their generalization for arbitrary $k$-quark motion was obtained in \cite{hartnoll}. They are characterized by the fact that the D5-brane wraps an $\bS^4\subset\bS^5$ at fixed polar angle $\vartheta\equiv\vartheta_1$, in the usual coordinatization where the 5-sphere metric reads
\begin{equation}\label{metricsphere}
G_{MN}dx^{M} dx^{N}=L^2(d\vartheta_1^2+\sin^2\vartheta_1 d\vartheta_2^2 +\ldots + \sin^2\vartheta_1\sin^2\vartheta_2\sin^2\vartheta_3\sin^2\vartheta_4 d\vartheta_5^2)~.
\end{equation}

If we are given a $k$-quark trajectory $x^{\mu}(\tau)$ and set out to construct a D5-brane of this type using our light ray method, by analogy with the D3-brane story we are led to pick $\xi^{\alpha}=(t_r,z,\vartheta_2,\vartheta_3,\vartheta_4,\vartheta_5)$ as a convenient choice of worldvolume coordinates. We are then required to shoot light inward from the AdS boundary, from each worldline point $x^{\mu}(t_r)$, along each of the directions on the $\bS^4$,  thereby tracing out null geodesics parametrized by $z$. Since we are specifically looking for embeddings with $\vartheta=$constant, these geodesics lie purely within AdS, and have no motion whatsoever along the $\bS^5$. But then they have no choice but to coincide with the by now familiar straight lines (\ref{mikhsolnc}), meaning that the AdS part of the D5-brane embedding is \emph{identical} to the fundamental string embedding dual to a quark with the same trajectory. And indeed, this is precisely the form of the embeddings deduced in \cite{hartnoll}.

To have a complete solution, we still need to determine the profile of the gauge field. Taking the hint from (\ref{Fg}), we propose a proportionality relation with the induced metric,
\begin{equation}\label{FgD5}
2\pi \ls^2 F_{\alpha\beta}=f g_{\alpha\beta}\qquad\forall\quad \alpha<\beta~,
\end{equation}
 with $f$ to be determined. Using (\ref{mikhsolnc}), this says that only the electric component $F_{t_r z}$ is nonvanishing, and it takes a value proportional to $g_{t_r z}=-L^2/\gamma z^2$. (Analogs of the simplifications (\ref{bisimp}) and (\ref{ngsimp}) are then in play.) This proposal indeed can be seen to satisfy the equation of motion for $\vartheta$, but only if the proportionality constant is fixed to $f=-\cos\vartheta$. We thus learn that
 \begin{equation}\label{FtzD5}
 2\pi \ls^2 F_{t_r z}=\frac{L^2\cos\vartheta}{\gamma z^2}~.
 \end{equation}
 All that remains is to enforce the condition that the fundamental string charge carried by the D5-brane be integer and equal to the required value, $\int d^4\vartheta\,\Pi^{t_r z}=k$. In other words, the solution must satisfy
 \begin{equation}\label{PitzD5}
 \Pi^{t_r z}=\frac{k\,\mbox{vol}(\bS^4)}{8\pi^2/3}~,
 \end{equation}
 with $\mbox{vol}(\bS^4)=\sin^3\vartheta_2\sin^2\vartheta_3\sin\vartheta_4$.
 (This is the direct analog of (\ref{Pitz}), and all other $\Pi^{ab}$ are found to correctly vanish.) This condition determines $\vartheta$ in terms of $k$,
 \begin{equation}\label{thetak}
 \sin\vartheta\cos\vartheta-\vartheta=\frac{\pi k}{N}~.
 \end{equation}
 Remarkably, (\ref{mikhsolnc}), (\ref{FtzD5}) and (\ref{thetak}) precisely agree with the solutions of \cite{pr,cpr,hartnoll}.

Now that we have obtained with our method both the D3-brane and D5-brane solutions respectively dual to the totally symmetric and antisymmetric representations of $SU(N)$, the same procedure is guaranteed to cover the case of $k$-quarks in \emph{arbitrary} irreducible representations. Indeed, it was shown in \cite{gp} that the dual of a source labelled by an arbitrary Young diagram can be understood as a collection of either D3-branes (if the diagram is decomposed into rows) or D5-branes (if it is decomposed into columns). At leading order in the large $N$ expansion, these branes do not interact with one another, and are therefore exactly as we have described in this section.

The fact that the D5-brane turns out to have the same structure as the D3-brane strengthens the hope that our light ray method will have more general applicability. For instance, we would expect it to also account successfully for the more general antisymmetric $k$-quark embeddings obtained in \cite{baryon}, of which the solutions described in this subsection are a special case. The situation there is more interesting because $\vartheta$ becomes a function of $z$, meaning that the relevant null geodesics will necessarily differ from (\ref{mikhsolnc}). We could similarly consider cases where $\vartheta$ depends in addition (or alternatively) on $t_r$.
The supersymmetric loops of \cite{zarembo,dgrt} are other notable examples associated with a variable direction on the $\bS^5$.
It is also interesting to note that the connection established in \cite{hartnoll} between D5-brane and  string embeddings does not require that the spacetime be pure AdS, which again suggests that the null geodesic construction could hold in more general backgrounds. We leave an exploration of these and other extensions to future work.

\section{Intrinsic and Radiative Energy}

As an application of our newly found D3-brane solutions, in this section we will compute their total energy and show that, similarly to \cite{mikhailov,dragtemp}, it cleanly splits into 2 contributions that represent the intrinsic energy of the $k$-quark and the energy that has already been carried away by gluonic (and scalar) radiation. This exercise will thus lead in particular to a holographic prediction for the rate of radiation of a(n infinitely massive) symmetric $k$-quark at strong coupling.

\subsection{D3-brane energy density}
We wish to compute the energy $E$ of the D3-brane at a fixed observation time $t$. For this purpose, unlike what we did in the previous section, it is convenient to choose as worldvolume coordinates $(t,z,\theta,\phi)$. As before, we use indices $\mu,\nu$ for the  Minkowski directions of the target AdS space, $\alpha,\beta$ for worldvolume coordinates, and $a,b$ for spatial worldvolume indices ($a,b=z,\theta,\phi$). Starting from the D3-brane action (\ref{D3action}), the energy density in our chosen parametrization is found to be \cite{fg}
\begin{equation}
{\cal E}=T_{D3}\left(
\frac{L^2}{z^2}\frac{\det(g_{ab}+2\pi \ls^2 F_{ab})}{\sqrt{-\det(g_{\alpha\beta}+2\pi \ls^2 F_{\alpha\beta})}}-
\frac{L^4}{z^4}\frac{\partial X^{\mu}}{\partial z} \frac{\partial X^{\nu}}{\partial \theta} \frac{\partial X^{\lambda}}{\partial \phi} \epsilon_{0\mu\nu\lambda}
\right)~.
\label{energydens}
\end{equation}
To be clear, the numerator in the first term is a $3\times 3$ determinant, with spatial indices only.

The full Born-Infeld determinant that appears in the denominator is available to us from Appendix~\ref{inducedapp}, albeit in the $(t_r,z,\theta, \phi)$ parametrization.  Applying to (\ref{bi}) a change of variables to our desired set $(t,z,\theta,\phi)$ yields
\begin{equation}
\det(g_{\alpha\beta}+2\pi\ls^2 F_{\alpha\beta})=-\frac{L^8}{z^4}\left(\frac{\partial t_r}{\partial t}\right)^2 \kappa^6 \gamma^{-2}\sin^2 \theta~.
\label{thebidet}
\end{equation}

Let us now compute the spatial determinant in the numerator of (\ref{energydens}). To do so, define $M=g+2\pi\ls^2F$ and notice that
$\det M_{ab}=M^{00}\det M_{\alpha\beta}$,
 where $M$ with indices above denotes the inverse matrix. Our strategy is to compute the matrix inverse to $M$ in the coordinates $(t_r,z,\theta,\phi)$, since in these coordinates the calculation is simpler, and then obtain $M^{00}$ in the $(t,z,\theta,\phi)$ coordinates by a change of variables.
This gives
\begin{eqnarray}
\left(\frac{\partial t}{\partial t_r}\right)^2 \det(g_{ab}+2\pi \ls^2 F_{ab})&=&
g_{\theta \theta} g_{\phi \phi} \partial_z t\left(g_{t_r t_r} \partial_z t-2g_{t_r z}\partial _{t_r}t\right)
 \\
{}&{}&
-\frac{\kappa^2}{1+\kappa^2}g_{\theta \theta}\left(g_{t_r\phi}\partial_zt-g_{t_r z}\partial_\phi t\right)^2
\nonumber\\
{}&{}&
-\frac{\kappa^2}{1+\kappa^2}g_{\phi \phi}\left(g_{t_r\theta}\partial_zt-g_{t_r z}\partial_\theta t\right)^2~.
\nonumber
\end{eqnarray}
A rather long computation then leads to\footnote{To arrive at the simplified expression (\ref{subdeter}) we have used the fact that any 4-vector
can be decomposed in terms of the tetrad (\ref{tetrad}), which implies for instance that
$\p_{t_r}v^{\mu}=(\p_{t_r}v\cdot n_1)n^{\mu}_1+(\p_{t_r}v\cdot n_2)n^{\mu}_2+(\p_{t_r}v\cdot n_3)n^{\mu}_3$.\label{tetradfoot}}
\begin{eqnarray} \label{subdeter}
\left(\frac{\partial t}{\partial t_r}\right)^2 \det(g_{ab}+2\pi \ls^2 F_{ab})=
\frac{L^6\kappa^4 \sin^2\theta}{z^2}\Bigg[(1+\kappa^2)-\kappa^2 \gamma^{-2}\left( (n_1^0)^2+(n_2^0)^2+(n_3^0)^2\right)
\nonumber\\
+  2z \partial_z t \gamma^{-1}\left( \left(\partial_{t_r}v^0+\frac{\kappa}{\sqrt{1+\kappa^2}}\partial_{t_r} n^0\right)+\gamma \frac{\kappa}{\sqrt{1+\kappa^2}}\partial_{t_r} n\cdot v\right)
+z^2(\partial_z t)^2 \left(\partial_{t_r}v \right)^2  \Bigg] .\quad
\end{eqnarray}
Note that the first line above contains terms independent of the acceleration, while the second line contains terms linear and quadratic in the acceleration.

The last piece in (\ref{energydens}) is the contribution from the Wess-Zumino term. We start by noticing that
\begin{equation}
\epsilon_{0\nu\lambda\rho} \partial_z X^\nu \partial_\theta X^\lambda \partial_\phi X^\rho=
\kappa^2 z^2\left(\frac{\partial t_r}{\partial t}\right) \sin\theta \;
\epsilon_{\mu \nu\lambda\rho} \partial_{t_r}X^\mu \partial_z X^\nu n_2^\lambda n_3^\rho~.
\end{equation}
Then we make use of property (\ref{tetradlevi}) to arrive at
\begin{equation}\label{wzterm}
\frac{\partial X^\mu}{\partial z} \frac{\partial X^\nu}{\partial \theta} \frac{\partial X^\lambda}{\partial \phi} \epsilon_{0 \mu \nu \lambda} =
\kappa^2 z^2 \left(\frac{\partial t_r}{\partial t}\right) \sin\theta \; (\kappa \gamma^{-1}+z\partial_{t_r}n_1\cdot v)~.
\end{equation}

Plugging (\ref{thebidet}), (\ref{subdeter}) and (\ref{wzterm}) into the expression for the energy density (\ref{energydens}), we find
\begin{eqnarray}\label{enerdens}
{\cal E}&=&\frac{N\kappa}{2\pi^2 z^2} \left(\frac{\partial t_r}{\partial t}\right) \sin \theta \, \Bigg[
\gamma +\kappa^2 \gamma^{-1}\left(\gamma^2-1-(n_1^0)^2-(n_2^0)^2-(n_3^0)^2\right)
\nonumber\\
{}&{}&\qquad\qquad
+2z \partial_z t \left( \left(\partial_{t_r}v^0+\frac{\kappa}{\sqrt{1+\kappa^2}}\partial_{t_r} n^0\right)+ \frac{\kappa}{\sqrt{1+\kappa^2}}\gamma \partial_{t_r} n\cdot v\right)-\kappa z\partial_{t_r} n\cdot v
\nonumber\\
{}&{}& \qquad\qquad+\gamma z^2 (\partial_z t)^2 \left(\partial_{t_r}v\right)^2 \Bigg].
\end{eqnarray}
The first, second and third line correspond respectively to terms whose explicit dependence on the acceleration is nonexistent, linear and quadratic (there is also implicit dependence in the $n_I$). In the first line there appears the combination
$\gamma^2-1-(n_1^0)^2-(n_2^0)^2-(n_3^0)^2$, which can be seen to vanish due to (\ref{fw}): differentiate with respect to $\tau$ to show that it is constant; then by boundary conditions it is zero.

\subsection{Total energy and rate of radiation}
We consider first the integral for terms that are independent of or linear in the acceleration,
\begin{eqnarray}\label{Ea}
E_{1}&=&\frac{N\kappa}{2\pi^2} \int dz d\Omega \, \frac{\partial_t t_r}{z^2}
\left[\gamma+2z\partial_z t\left(\left(\partial_{t_r}v^0+\frac{\kappa}{\sqrt{1+\kappa^2}}\partial_{t_r} n^0\right)
\right.\right.
\\
{}&{}&\qquad\qquad\qquad\qquad\qquad\qquad\qquad
+\frac{\kappa}{\sqrt{1+\kappa^2}}\gamma \partial_{t_r} n\cdot v\bigg)-\kappa z\partial_{t_r} n\cdot v\bigg]~,
\nonumber
\end{eqnarray}
with $d\Omega=\sin\theta d\theta d\phi$.
It is convenient to change the integration variable $z\rightarrow t_r$, using
\begin{equation}\label{ztotr}
\frac{\partial z}{\partial t_r}=-\frac{1}{(\partial_t t_r)( \partial_z t)}~.
\end{equation}
This cancels the $\partial_t t_r$ prefactor in (\ref{Ea}). We must also use (\ref{D3covariant}) to substitute $z$ in terms of $t_r$ in the integrand, but for compactness we keep it as a shorthand.

In the string analysis of \cite{mikhailov}, at this stage a total derivative, $d(\gamma z^{-1})/dt_r$, was discarded from the integrand, which was later reinstated in \cite{dragtemp} and shown to have physical significance: upon integration, it yields the intrinsic energy of the quark at time $t$. Inspired by this, we observe that in our D3 context too the derivative
\begin{equation}\label{totalder}
\frac{\p (\gamma z^{-1})}{\p t_r}=\frac{\gamma}{z^2\partial_z t}+\frac{\p_{t_r}\gamma}{z}+\frac{\sqrt{1+\kappa^2}}{z \partial_z t}\gamma\left(\partial_{t_r}v^0+\frac{\kappa}{\sqrt{1+\kappa^2}}\partial_{t_r} n^0\right)~,
\end{equation}
 is similar to the integrand we have, but differs by terms proportional to $n^{\mu}$:
 \begin{equation}\label{Ea2}
E_{1}=\frac{N\kappa}{2\pi^2} \int dt_r d\Omega \,
\left[\frac{\p (\gamma z^{-1})}{\p t_r}
+\frac{\kappa}{t-t_r}\left(n^0\p_{t_r}\gamma-(\gamma^2-1)n\cdot\p_{t_r}v\right)\right]~.
\end{equation}
Remembering that $n^{\mu}$ is obtained from its rest frame counterpart $n^{\mu}_R$ by a boost followed by Fermi-Walker transport (\ref{fw}), we see that each of its components is linear in the components of $n^{\mu}_R$ (with coefficients that depend on the 4-velocity and 4-acceleration), so their angular integral at fixed $t_r$ vanishes,
\begin{equation}\label{nintegral}
\int d\Omega \;  n^{\mu}_R \; = 0 \quad {\Longrightarrow} \quad \int d\Omega \; n^{\mu}=0~.
\end{equation}

For the remaining, total derivative term, it is evidently convenient to carry out the $t_r$ integral first, to be left with just a surface term at $t_r=t$. The result is in fact divergent, and we choose to regularize it by cutting off the D3-brane at a fixed radial position $z_{\mbox{\scriptsize min}}$ (so the cutoff in $t_r$
depends on $\theta,\phi$).\footnote{Alternatively, one can choose to cut off the integral at $t_r=t-\epsilon$, with $\epsilon$ constant, in which case the corresponding $z_{\mbox{\scriptsize min}}$ would depend on $\theta,\phi$. The results of the two regularizations agree in the limit where the cutoff is removed.}
 The leading contribution as the cutoff is removed is then just $\gamma(t)/z_{\mbox{\scriptsize min}}$, which is spherically symmetric, so the angular integral is trivial. We are thus left with
\begin{equation}\label{Easimp}
E_{1}=\lim_{z_{\mbox{\tiny min}}\to 0}\frac{2N\kappa\gamma(t)}{\pi z_{\mbox{\scriptsize min}}}=k m\gamma(t)~,
\end{equation}
where $m=\sqrt{\lambda}/2\pi z_{\mbox{\scriptsize min}}$ denotes the rest mass of a quark ($k=1$) with the same UV cutoff.

Consider now the integral of the terms quadratic in the acceleration. After the change of variable $z\rightarrow t_r$ it can be rewritten as
\begin{equation} \label{Easq}
E_{2}=\frac{N\kappa}{2\pi^2}\int dt_r d\Omega \, \gamma\p_z t  \, (\partial_{t_r} v)^2~.
\end{equation}
Performing the integral over $\theta,\phi$ and using (\ref{nintegral}) again, we finally obtain
\begin{equation}\label{Easqsimp}
E_{2}=\frac{2N}{\pi}\kappa\sqrt{1+\kappa^2}\int dt_r\,\gamma^2 \, (\partial_{t_r} v)^2
=\frac{2N}{\pi}\kappa\sqrt{1+\kappa^2}\int dt_r\,\gamma^6(\vec a^2-|\vec v\times \vec a|^2)~.
\end{equation}

As promised, we have thus found by explicit computation that, much as in \cite{mikhailov,dragtemp}, the total energy of the D3-brane can be understood as the sum of 2 contributions, (\ref{Easimp}) and (\ref{Easqsimp}), that admit a pleasant and direct gauge-theoretic interpretation. $E_1$, which depends only on the state of motion of the probe at the observation time $t$, is the energy attributable to the $k$-quark itself, while $E_2$, which depends on the entire previous history, is the total energy that has been radiated away. The fact that these quantities can be determined analytically in the strong coupling regime constitutes yet another illustration of the power of the AdS/CFT correspondence.

{}From (\ref{Easimp}), which agrees with the result (\ref{intenergysym}) that we had advertised in the Introduction, we see that the $k$-quark has an intrinsic energy that, even for arbitrary motion, equals that of $k$ individual quarks, i.e., it is a threshold bound state. This is a direct consequence of its pointlike nature, Lorentz invariance, and the fact that it must obey the BPS bound.\footnote{It is worth noting that, as the UV cutoff is removed to obtain (\ref{Easimp}), there is a subleading, order $\epsilon^0$ correction to the intrinsic energy of both the quark and the $k$-quark. This contribution, proportional to $\vec{v}\cdot\vec{a}$, is closely related to the results of \cite{dragtemp,lorentzdirac,damping} for a finitely-massive quark, and also, to the surprising fact that the `near field' of the probe has a tail that reaches out to infinity \cite{iancuradiation,tmunu}.}

Expression (\ref{Easqsimp}), which we had advertised in (\ref{radenergysym}), amounts to a prediction for the rate of radiation of a totally symmetric color source in the strongly-coupled gauge theory: it is given by the usual Li\'enard formula, with a numerical coefficient that matches the findings of \cite{fg,fgl} in the context of a more indirect calculation which examined only the specific case of uniform acceleration. Now that we have access to the D3-brane dual to a symmetric $k$-quark undergoing arbitrary motion, we are able to verify that the functional form of the rate of energy loss is unmodified, and the corresponding Bremsstrahlung function is indeed given by (\ref{bksym}) in the supergravity limit. For $k=1$, and at leading order in the $1/N$ expansion, (\ref{Easqsimp}) reproduces Mikhailov's result (\ref{radenergyfund}) for the rate of radiation of a quark. And, since the former coincides with the result of \cite{fg,fgl}, it shares the remarkable property that the entire series of corrections in powers of $\kappa\propto \sqrt{\lambda}/N$  matches onto the exact result obtained in \cite{correa,lewkomalda} for a probe in the fundamental representation, in the limit where $N,\lambda\to\infty$ with $\sqrt{\lambda}/N$ fixed.

\section*{Acknowledgements}
We are grateful to Mariano Chernicoff, Roberto Emparan, Jaume Garriga, Simone Giombi, Veronika Hubeny, Igor Klebanov, Aitor Lewkowycz and David Mateos for useful discussions, and to Mariano Chernicoff for suggestions on the manuscript. BF would like to thank the organizers of Mextrings 2014 for the welcoming atmosphere that ignited this collaboration. He is also grateful to the Crete Center for Theoretical Physics, the Theory Division at CERN and also CERN-Korea Theory Collaboration funded by National Research Foundation (Korea),  for hospitality during the course of this work. The research of BF  is supported by MEC FPA2010-20807-C02-02, CPAN CSD2007-00042, within the Consolider-Ingenio2010 program, and AGAUR 2009SGR00168.
AG is partially supported by Mexico's National Council of Science and Technology (CONACyT) grant 104649, DGAPA-UNAM grant IN110312, and sabbatical fellowships from DGAPA-UNAM and CONACyT. He would also like to thank the Department of Physics of Princeton University, and Igor Klebanov in particular, for hosting his sabbatical.
JFP is partially supported by the National Science Foundation under Grant No. PHY-1316033 and by Perimeter Institute for Theoretical Physics. Research at Perimeter Institute is funded by the Government of Canada through Industry Canada and by the Province of Ontario through the Ministry of Research and Innovation.

\appendix

\section{Completing the embedding for uniform acceleration}\label{accelapp}

For the case of a source in the fundamental representation undergoing uniform acceleration $A$, it was found in \cite{brownian,noline} that the embedding provided by Mikhailov is in fact incomplete. Instead of extending all the way up to the Poincar\'e horizon at $z\to\infty$, it terminates at the radial position $z_{h}=A^{-1}$, which marks the location of a worldsheet horizon. This can be seen directly from the first equation in (\ref{mikhsolnc}): if we consider the embedding at fixed $t$, that equation would normally imply that $z\to\infty$ as $t_r\to-\infty$. But precisely for uniform acceleration, the factor of $\gamma$ diverges linearly in $t_r$ in that limit, so $z$ remains finite. This is not a mistake: the resulting portion of the string truly codifies the gluonic field sourced by the quark since the beginning of time.
But the string evidently cannot end in midair, so it ought to be continued somehow. Its continuation encodes the initial conditions of the gluonic field at $t_r\to-\infty$.

There are three options for this continuation.\footnote{We thank Veronika Hubeny for extensive discussions on this point.} If one insists on achieving a smooth worldsheet and preserving the purely retarded structure of the field sourced by the quark, then one is forced to include a mirror antiquark \cite{noline,hs}, i.e., the string must be continued back to the AdS boundary, by pasting it together with an appropriate purely advanced solution. This yields the semicircular  quark-antiquark embedding found independently in \cite{xiao} (see also \cite{ppz}). (In fact, this same pasting method can be used to construct an infinite family of smooth solutions where the quark and antiquark are only required to approach uniform acceleration in the remote past and, respectively, future \cite{eprer}.) If one preserves the retarded structure and insists on having no antiquark, then the embedding cannot be smooth, and the missing portion of the string is found to be a purely radial segment moving at the speed of light \cite{noline}, which codifies a gluonic shock wave that is present already in the asymptotic past and is progressively shed by the quark. Finally, if one insists on a smooth embedding and on having an isolated quark, then one must give up the retarded structure and allow incoming waves from infinity \cite{hs}.

When we move on to the totally symmetric $k$-quark case, the situation is entirely analogous. {}From the first equation in (\ref{D31D})
 we see that the D3-brane embedding also terminates at a finite radial location $z_h$ in AdS, which depends on the polar angle $\theta$  according to $z_{h}=A^{-1}/(\sqrt{1+\kappa^2}-\kappa\cos\theta)$. It is easy to check that at this location $g_{t_r t_r}=0$, i.e., we again have a worldvolume horizon. The second equation in (\ref{D31D}) implies that this termination locus is given by $x^1=-t_r$, and thus translates in the negative $x^1$ direction at the speed of light.
 The banana-shaped embedding obtained in \cite{df,fg} and displayed in Fig.~\ref{afig} is then understood to be analogous to the quark-antiquark solution  of \cite{xiao}: it is obtained by pasting together the retarded solution (\ref{D31D}) for a $k$-quark with an advanced solution (i.e., a time-inverted copy of (\ref{D31D})) for a $k$-antiquark with the same acceleration. (Again, as in \cite{eprer}, more general solutions can be obtained where the sources are only asymptotically uniformly accelerated.) If we insist on having a purely retarded solution describing just a quark and no antiquark, we would need to complete the embedding differently, with a vertical portion moving at the speed of light, as in \cite{noline}. Or if we prefer to stick with smooth embeddings for the case of an isolated quark, then we must forego the purely retarded (advanced) structure, as in \cite{hs}.

 As explained in Section \ref{d5subsec}, the case of a totally antisymmetric color source is described by a D5-brane whose AdS dynamics exactly coincide with those of a string with the same trajectory at the boundary \cite{hartnoll}, and will therefore also allow only these same 3 types of completions.

 \section{Induced metric and Born-Infeld determinant}\label{inducedapp}

In this Appendix we will work out the induced metric and Born-Infeld determinant for the D3-brane embedding (\ref{D3covariant}), with $(t_r,z,\theta,\phi)$ as worldvolume coordinates. The resulting expressions are needed at a couple of places in the main text.

Start by noticing that
\begin{equation}
\p_z X^{\mu}\rvert_{t_r,\theta,\phi}=\sqrt{1+\kappa^2}\, v^\mu+\kappa n^\mu~,
\end{equation}
so using $v^2=-1, v\cdot n=0, n^2=1$ it follows that $\p_z X^\mu \p_z X_\mu=-1$, and therefore
\begin{equation}\label{gzz}
g_{zz}=0~.
\end{equation}
As stated in Section \ref{d3subsec}, this embodies the requirement that the lines at fixed $t_r,\theta,\phi$ be null.
{}The properties of $v^{\mu}$ and $n^{\mu}$ additionally imply that
$$
 v\cdot \partial_\theta n= v\cdot \partial_\phi n= n\cdot \partial_\theta n= n\cdot \partial_\phi n=0~,
$$
and from this and (\ref{D3covariant}) it follows that
\begin{equation}\label{gztheta}
g_{z\theta}=g_{z\phi}=0~.
\end{equation}

For the remaining spatial components, we will use the fact that the initial boost (\ref{boost}) and subsequent Fermi-Walker transport (\ref{fw}) that relates $n^{\mu}$ to the rest frame vector $n^{\mu}_R$ do not depend themselves on $\theta,\phi$. Consequently, the relations $\partial_\theta n_R \cdot \partial_\phi n_R=0$, $\partial_\theta n_R \cdot \partial_\theta n_R=1$, $\partial_\phi n_R \cdot \partial_\phi n_R=\sin^2 \theta$ imply the corresponding properties
$$
\partial_\theta n \cdot \partial_\phi n=0~,\qquad
\partial_\theta n \cdot \partial_\theta n=1~,\qquad
\partial_\phi n \cdot \partial_\phi n=\sin^2 \theta~,
$$
which combined with (\ref{D3covariant}) lead in turn to
\begin{equation}\label{gthetaphi}
g_{\theta \phi}=0~,\qquad g_{\theta \theta}=L^2 \kappa^2~,\qquad  g_{\phi \phi}=L^2 \kappa^2 \sin^2 \theta~.
\end{equation}

For the metric components with a $t_r$ index, we find
\begin{eqnarray}\label{gtr}
g_{t_r t_r}&=&\frac{L^2}{z^2}\left(-\gamma^{-2}+2\kappa z\gamma^{-1}v\cdot \partial_{t_r} n_1+z^2\left(\sqrt{1+\kappa^2}\partial_{t_r}v+\kappa \partial_{t_r}n_1\right)^2 \right)~,
\nonumber\\
g_{t_r z}&=&-\frac{L^2}{z^2}\sqrt{1+\kappa^2}\gamma^{-1}~,
\\
g_{t_r \theta}&=&L^2 \kappa \sqrt{1+\kappa^2}\partial_{t_r}v \cdot n_2~,
\nonumber\\
g_{t_r \phi}&=&L^2 \kappa \sin \theta \sqrt{1+\kappa^2}\partial_{t_r}v \cdot n_3~.
\nonumber
\end{eqnarray}
In the last 2 equations, we used the definition of the $n_I$ given above (\ref{tetrad}).

With these results, it is easy to verify that the determinant of the metric indeed simplifies to (\ref{ngsimp}).

Next, we compute the Born-Infeld determinant. From the form of the induced metric and the fact that in these coordinates the Born-Infeld field is purely electric, it follows immediately that
$$
\det(g_{\alpha\beta}+2\pi\ls^2 F_{\alpha\beta})=-\left(g_{t_rz}^2-(2\pi \ls^2 F_{t_rz})^2\right)g_{\theta \theta}g_{\phi \phi}=-\frac{\kappa^2}{1+\kappa^2}g_{t_rz}^2 g_{\theta \theta}g_{\phi \phi}~,
$$
where in the last step we used (for the first time in this argument) the relation (\ref{Fg}) between $F$ and $g$. Therefore we arrive at
\begin{equation}\label{bi}
\det(g_{\alpha\beta}+2\pi\ls^2 F_{\alpha\beta})=-\frac{L^8}{z^4}\kappa^6 \gamma^{-2}\sin^2 \theta~.
\end{equation}
Comparing with (\ref{ngsimp}), we see that the relation (\ref{bisimp}) between the Dirac and Born-Infeld determinants indeed holds.

\end{document}